\documentclass[preprint]{aastex62}

\newcommand\lsun{L$_\sun$}
\newcommand\mearth{M$_\earth$}
\newcommand\ms{m s$^{-1}$}

\newcommand\msun{M$_\sun$}
\newcommand\prot{$P_{\rm rot}$}
\newcommand\rearth{R$_\earth$}
\newcommand\rsun{R$_\sun$}
\newcommand\searth{S$_\earth$}

\received{August 2, 2018}
\revised{September 6, 2018}
\accepted{November 15, 2018}
\submitjournal{AJ}

\shorttitle{A second terrestrial planet orbiting LHS 1140}
\shortauthors{Ment et al.}

\begin{document}
\title{A second terrestrial planet orbiting the nearby M dwarf LHS 1140}

\correspondingauthor{Kristo Ment}
\email{kristo.ment@cfa.harvard.edu}

\author[0000-0001-5847-9147]{Kristo Ment}
\affil{Harvard-Smithsonian Center for Astrophysics, 60 Garden Street, Cambridge, MA 02138, USA}

\author[0000-0001-7730-2240]{Jason A. Dittmann}
\affiliation{51 Pegasi b Postdoctoral Fellow, Massachusetts Institute of Technology, 77 Massachusetts Avenue, Cambridge, MA 02139, USA}

\author[0000-0002-8462-515X]{Nicola Astudillo-Defru}
\affil{Departamento de Astronom\'ia, Universidad de Concepci\'on, Casilla 160-C, Concepci\'on, Chile}

\author{David Charbonneau}
\affil{Harvard-Smithsonian Center for Astrophysics, 60 Garden Street, Cambridge, MA 02138, USA}

\author{Jonathan Irwin}
\affil{Harvard-Smithsonian Center for Astrophysics, 60 Garden Street, Cambridge, MA 02138, USA}

\author{Xavier Bonfils}
\affil{Universit{\'e} Grenoble Alpes, CNRS, IPAG, F-38000 Grenoble, France}

\author{Felipe Murgas}
\affil{Instituto de Astrof{\'i}sica de Canarias (IAC), E-38200 La Laguna, Tenerife, Spain\label{iac}}

\author{Jose-Manuel Almenara}
\affil{Universit{\'e} Grenoble Alpes, CNRS, IPAG, F-38000 Grenoble, France}

\author{Thierry Forveille}
\affil{Universit{\'e} Grenoble Alpes, CNRS, IPAG, F-38000 Grenoble, France}

\author{Eric Agol}
\affil{Astronomy Department, University of Washington, Seattle, WA 98195, USA}

\author{Sarah Ballard}
\affiliation{Massachusetts Institute of Technology, 77 Massachusetts Avenue, Cambridge, MA 02138, USA}

\author{Zachory K. Berta-Thompson}
\affil{Department of Astrophysical and Planetary Sciences, University of Colorado, Boulder, CO 80309, USA}

\author{Fran\c{c}ois Bouchy}
\affil{Observatoire de l'Universit{\'e} de Gen{\`e}ve, 51 chemin des Maillettes, 1290 Versoix, Switzerland}

\author{Ryan Cloutier}
\affil{Dept. of Astronomy \& Astrophysics, University of Toronto, 50 St. George Street, M5S 3H4, Toronto, ON, Canada}
\affil{Centre for Planetary Sciences, Dept. of Physical \& Environmental Sciences, University of Toronto Scarborough, 1265 Military Trail, M1C 1A4, Toronto, ON, Canada}
\affil{Institut de Recherche sur les Exoplan{\`e}tes, D{\'e}partement de physique, Universit{\'e} de Montr{\'e}al, CP 6128 Succ. Centre-ville, H3C 3J7, Montr{\'e}al, QC, Canada}

\author{Xavier Delfosse}
\affil{Universit{\'e} Grenoble Alpes, CNRS, IPAG, F-38000 Grenoble, France}

\author{Ren{\'e} Doyon}
\affil{Institut de Recherche sur les Exoplan{\`e}tes, D{\'e}partement de physique, Universit{\'e} de Montr{\'e}al, CP 6128 Succ. Centre-ville, H3C 3J7, Montr{\'e}al, QC, Canada}

\author{Courtney D. Dressing}
\affil{Department of Astronomy, University of California, Berkeley, CA 94720, USA}

\author{Gilbert A. Esquerdo}
\affil{Harvard-Smithsonian Center for Astrophysics, 60 Garden Street, Cambridge, MA 02138, USA}

\author{Rapha{\"e}lle D. Haywood}
\affil{Harvard-Smithsonian Center for Astrophysics, 60 Garden Street, Cambridge, MA 02138, USA}

\author{David M. Kipping}
\affil{Department of Astronomy, Columbia University, 550 W 120th Street, New York, NY 10027, USA}

\author{David W. Latham}
\affil{Harvard-Smithsonian Center for Astrophysics, 60 Garden Street, Cambridge, MA 02138, USA}

\author{Christophe Lovis}
\affil{Observatoire de l'Universit{\'e} de Gen{\`e}ve, 51 chemin des Maillettes, 1290 Versoix, Switzerland}

\author{Elisabeth R. Newton}
\affiliation{Massachusetts Institute of Technology, 77 Massachusetts Avenue, Cambridge, MA 02138, USA}

\author{Francesco Pepe}
\affil{Observatoire de l'Universit{\'e} de Gen{\`e}ve, 51 chemin des Maillettes, 1290 Versoix, Switzerland}

\author{Joseph E. Rodriguez}
\affil{Harvard-Smithsonian Center for Astrophysics, 60 Garden Street, Cambridge, MA 02138, USA}

\author{Nuno C. Santos}
\affil{Instituto de Astrof{\'i}sica e Ci{\^e}ncias do Espa\c{c}o, Universidade do Porto, CAUP, Rua das Estrelas, 4150-762 Porto, Portugal}
\affil{Departamento de F{\'i}sica e Astronomia, Faculdade de Ci{\^e}ncias, Universidade do Porto, Rua do Campo Alegre, P-4169-007 Porto, Portugal}

\author{Thiam-Guan Tan}
\affil{Perth Exoplanet Survey Telescope, Perth, Western Australia 6010, Australia}

\author{Stephane Udry}
\affil{Observatoire de l'Universit{\'e} de Gen{\`e}ve, 51 chemin des Maillettes, 1290 Versoix, Switzerland}

\author{Jennifer G. Winters}
\affil{Harvard-Smithsonian Center for Astrophysics, 60 Garden Street, Cambridge, MA 02138, USA}

\author{Ana{\"e}l W{\"u}nsche}
\affil{Universit{\'e} Grenoble Alpes, CNRS, IPAG, F-38000 Grenoble, France}

\begin{abstract}

LHS 1140 is a nearby mid-M dwarf known to host a temperate rocky super-Earth (LHS 1140 b) on a 24.737-day orbit. Based on photometric observations by MEarth and Spitzer as well as Doppler spectroscopy from HARPS, we report the discovery of an additional transiting rocky companion (LHS 1140 c) with a mass of 1.81 $\pm$ 0.39 \mearth\ and a radius of 1.282 $\pm$ 0.024 \rearth\ on a tighter, 3.77795-day orbit. We also obtain more precise estimates of the mass and radius of LHS 1140 b to be 6.98 $\pm$ 0.89 \mearth\ and 1.727 $\pm$ 0.032 \rearth. The mean densities of planets b and c are $7.5\pm1.0~\rm{g/cm^3}$ and $4.7\pm1.1~\rm{g/cm^3}$, respectively, both consistent with the Earth's ratio of iron to magnesium silicate. The orbital eccentricities of LHS 1140 b and c are consistent with circular orbits and constrained to be below 0.06 and 0.31, respectively, with 90\% confidence. Because the orbits of the two planets are co-planar and because we know from previous analyses of \emph{Kepler} data that compact systems of small planets orbiting M dwarfs are  commonplace, a search for more transiting planets in the LHS 1140 system could be fruitful. LHS 1140 c is one of the few known nearby terrestrial planets whose atmosphere could be studied with the upcoming \emph{James Webb Space Telescope}.

\end{abstract}

\keywords{planets and satellites: detection, terrestrial planets --- techniques: photometric, radial velocities}

\section{Introduction} \label{sec:intro}

Small planets are very common around M dwarfs \citep{Dressing2013,Bonfils2013,Mulders2015}, with a cumulative occurrence rate of at least 2.5 $\pm$ 0.2 planets per M dwarf \citep{Dressing2015a}. Compared to Sun-like counterparts, such planets are easier to detect due to the smaller sizes and lower masses of the stars, which translate into larger transit depths and Doppler signals. Moreover, the detection and follow-up studies of terrestrial planets in the habitable zone are facilitated by the fact that the same stellar insolation as that received by the Earth is achieved at shorter orbital periods around M dwarfs. The increased frequency and greater depth of transits also renders the atmospheres of these worlds more accessible to transmission and emission spectroscopy, and atmospheric studies are eagerly anticipated with both the upcoming \emph{James Webb Space Telescope} \citep[e.g.][]{Morley2017} and the Extremely Large Ground-based Telescopes \citep[e.g.][]{Rodler2014}. For these reasons, the only terrestrial exoplanets whose atmospheres can be spectroscopically studied in the near-future will be those that orbit nearby mid-to-late M dwarfs. Considering stars with masses less than $0.3 M_{\sun}$ and within 15 parsecs, there are only four such stars known to host transiting planets: GJ 1214 \citep{Charbonneau2009}, GJ 1132 \citep{BertaThompson2015}, TRAPPIST-1 \citep{Gillon2017}, and LHS 1140 \citep{Dittmann2017}. There are also an additional 7 non-transiting systems: Proxima \citep{Anglada-Escude2016}, Ross 128 \citep{Bonfils2018Ross128}, YZ Cet \citep{Astudillo-Defru2017b}, GJ 273 and GJ 3323 \citep{Astudillo-Defru2017a}, Wolf 1061 \citep{Astudillo-Defru2017a,Wright2016}, and Kapteyn's star \citep{Anglada-Escude2014}.

The \emph{Kepler dichotomy} refers to the observed excess of stars with single transiting planets compared to the expectations based on geometry and the population of stars with multiple transiting planets \citep{Lissauer2011}. \cite{Ballard2016} studied this effect for M dwarfs, and found they could account for the observed Kepler population if they posited that half of the planetary systems have on average 7 planets with small mutual inclinations, while the other half of planetary systems have only a single close-in planet, or multiple planets with a large range of mutual inclinations. Numerous compact systems of small planets orbiting M dwarfs are known. In fact, \cite{Muirhead2015} found that 21$^{+7}_{-5}$\% of the Kepler M dwarfs host multiple planets with orbital periods less than 10 days. Indeed, of the 11 planetary systems of nearby mid-to-late M dwarfs described in the preceding paragraph, the majority have (or are suspected to have) more than one planet. Our intensive campaign with HARPS to follow-up GJ 1132 recently revealed the presence of at least one additional planet \citep{Bonfils2018}, although it does not appear to transit \citep{Dittmann_1132}.

\citet{Dittmann2017} announced the discovery of a terrestrial planet on a 24.7-day orbit in the habitable zone of LHS 1140, with an estimated zero-albedo equilibrium temperature of 230 $\pm$ 20 K. Given the context above, LHS 1140 is a particularly attractive target to search for co-planar planets orbiting interior to the known planet. Therefore, since the discovery of LHS 1140 b, we have intensively monitored this star with MEarth and HARPS in the hope that we would uncover additional transiting planets. A secondary science goal of these observations is to refine the estimate of the density of LHS 1140 b, and hence to address whether the ratio of the iron core to the rocky mantle is consistent with that of the Earth and terrestrial planets generally, given the small spread in abundances of Fe, Mg, and Si among nearby (albeit Sun-like) stars \citep{Bedell2018}. LHS 1140 is a slowly rotating and inactive star, and hence there is no reason to expect that stellar photospheric effects will set a limit to the precision with which the planetary mass may be determined. In this paper, we present a substantially improved estimate for the density of LHS 1140 b, and report the discovery of another small terrestrial planet, LHS 1140 c, on a 3.8-day orbit.

This paper is structured as follows. Section \ref{sec:star} summarizes our knowledge of the host star LHS 1140. Section \ref{sec:data} provides an overview of the different instruments and surveys that were used to conduct photometric and spectroscopic monitoring of LHS 1140. Section \ref{sec:discovery} outlines the process that led to the discovery of the new terrestrial companion LHS 1140 c. Subsequently, we performed joint modeling of the two planets to estimate their orbital parameters as described in Section \ref{sec:modeling}. We conclude by discussing the scientific implications of our findings in Section \ref{sec:discussion}.

\section{LHS 1140} \label{sec:star}

LHS 1140 is a M4.5-type main-sequence red dwarf \citep{Dittmann2017}. The most up-to-date parallax estimate for LHS 1140 is $\pi = 0.066700 \pm 0.000067\arcsec$ \citep{Gaia2018} which translates to a distance of 14.993 $\pm$ 0.015 pc. We note that this distance is substantially larger than the previous best estimate of 12.47 pc by \citet{Dittmann2017}. Using the mass-luminosity relation (MLR) for main-sequence M dwarfs by \citet{Benedict2016} and the 2MASS K-band apparent brightness $K_S = 8.821 \pm 0.024$, we obtain an estimate for the stellar mass of $M_\star = 0.179 \pm 0.014$ \msun. We calculated the uncertainty on the mass by propagating the errors in the parallax and $K_S$ values and adding a 0.014\msun\ term in quadrature to account for the scatter in the MLR. Therefore, we find that the uncertainty on the stellar mass is completely dominated by the scatter in the MLR.

We obtained an updated estimate for the radius of LHS 1140 from our transit models in Section \ref{sec:jointfit}. In particular, the $a/R_\star$ ratio (where $a$ is the orbital semi-major axis and $R_\star$ is the radius of the star) can be obtained directly from the transit duration, and the semi-major axis $a$ can be calculated using Kepler's Third Law knowing the orbital period $P$ and stellar mass $M_\star$. This also leads to a direct constraint on the stellar density from the transit parameters alone via $\rho_\star \approx \frac{3\pi}{GP^2} \left( \frac{a}{R_\star} \right)^3$, assuming a circular orbit \citep{Seager2003}. In particular, we obtain $R_\star = 0.2111 \pm 0.0059$ \rsun\ from the transit models for LHS 1140 b, and $R_\star = 0.2165 \pm 0.0057$ \rsun\ from the models for LHS 1140 c. We average the two and use $R_\star = 0.2139 \pm 0.0041$ as the final estimate. This result is also consistent with the mass-radius relation from long-baseline optical interferometry of single stars \citep{Boyajian2012} which yields $R_\star = 0.209 \pm 0.011$ \rsun.

Due to the change in the distance estimate to LHS 1140, the stellar luminosity $L_\star$ also needs to be reevaluated. We obtained a new value for the stellar luminosity by combining several bolometric correction estimates. In particular, the \citet{Leggett2001} relation between $BC_J$ and $I-K$ yields $L_\star$ = 0.00434 \lsun\ whereas the \citet{Mann2015} relation between $BC_K$ and $V-J$ gives $L_\star$ = 0.00431 \lsun. Alternatively, we can interpolate the $BC_V$ sequence of \citet{Pecaut2013} based on the $V-K$ color of the star to get $L_\star$ = 0.00459 \lsun. We adopt as our final estimate the mean and standard deviation of the three estimates, obtaining $L_\star$ = 0.00441 $\pm$ 0.00013 \lsun. We also adopt the solar values of \lsun = $(3.8270 \pm 0.0014) \cdot 10^{26}$ W and $M_{bol,\sun} = 4.7554$ used by \citet{Pecaut2013}. As a result, we can estimate the effective surface temperature from the Stefan-Boltzmann law as $T_{\rm eff,\star} = 3216 \pm 39$ K.

\citet{Dittmann2017} used the near-infrared spectral features of LHS 1140 to establish the metallicity of the star as [Fe/H] = -0.24 $\pm$ 0.10. They infer the age of LHS 1140 to be larger than 5 Gyr based on the lack of active H$\rm \alpha$ emission as well as the slow rotation of the star. We adopt these values.

The parameters of LHS 1140 can be found in Table \ref{tbl:results}.

\section{Photometric and Radial Velocity Data} \label{sec:data}

\subsection{MEarth} \label{sec:mearth}

MEarth-South is a telescope array consisting of eight 40-cm telescopes at the Cerro Tololo International Observatory (CTIO) in Chile. The telescopes are operated on a (nearly) automated basis and take data on every clear night. The observational strategy and the data reduction process are described in greater detail in \citep{jonathan_cool_stars} and \citet{Dittmann_1132}. We note that the MEarth data are corrected for the effect of differential color extinction by the Earth's atmosphere. The primary driver of differential color extinction in our data is the variability in the amount of precipitable water vapor in the atmosphere over the course of a night. Because the MEarth targets are mid-to-late M dwarfs and typically the reddest object in any observing field, our target stars are more sensitive to changes in water vapor than the field reference stars. We correct this effect by measuring a ``common mode" for all of our target M dwarfs. The common mode is defined as the average differential (i.e. relative flux) light curve of all M dwarfs currently being observed by MEarth-South in time bins of 0.02 days (28.8 minutes). This correlated behavior serves as a good proxy for the local change in precipitable water vapor on these time scales.

MEarth data are reduced in real time during the night. During the course of normal operations, MEarth-South is able to identify potential transits in-progress and instead of proceeding to the next star in its target list, it can automatically decide to halt the data collection of other targets in order to collect additional high-cadence data around the star showing signs of a potential transit event. Normal operations are then resumed if the event is deemed to be spurious or the flux from the star has returned to its normal level. This MEarth ``trigger" mode enables us to potentially confirm a planetary transit in real time and follow the transit all the way through egress. For a more detailed description of the MEarth trigger mode, see \citet{berta2012}.

MEarth-South has been in operation since January of 2014, and we have been observing LHS 1140 since the beginning of this survey. Since 2014 we have taken 28,382 observations of LHS 1140 with MEarth-South. The observations contain the discovery data of LHS 1140 b \citep{Dittmann2017}, including the high-cadence follow-up observations of LHS 1140 b's transits. The majority of these data, however, are standard monitoring observations of LHS 1140. Since October of 2015, the monitoring of LHS 1140 is carried out with two telescopes, with several exposures per visit.

\subsection{Spitzer} \label{sec:spitzer}

We observed four transits of LHS 1140 b as part of the $Spitzer$ DDT program 13174 (PI: Dittmann). These data were taken with the Infrared Array Camera (IRAC) at 4.5 $\mu$m. The goal of this program is to better determine the parameters of LHS 1140 b and to probe the system for signs of a transiting exomoon or transit timing variations (TTVs). A manuscript analyzing these 4 transits is currently under preparation. Here, we present one of these transit observations, which fortuitously captured a transit of both LHS 1140 b and LHS 1140 c. 

Observations spanned from 18 April 2018 03:21:22 to 18 April 2018 09:27:59 UTC. We placed LHS 1140 in the portion of the detector that is well-characterized for the purpose of obtaining high precision light curves. We obtained the data in subarray mode and utilized a small 16$\times$16 pixel area of the detector for our observation. During the first 78 minutes of the observations, the centroid of LHS 1140 wandered nearly 0.3 pixels (0.37") before settling into the ``sweet-spot" for the rest of the observations. In order to correct for intra-pixel sensitivity variations, we calibrate our data with the pixel-level decorrelation algorithm which depends on stable pointing to be effective. We exclude this portion of the data and only use data for which the target has settled onto the same portion of the detector. Subsequently, 50\% of the image centroids fall within 0.069 pixels of the median centroid location and 95\% of the image centroids fall within 0.150 pixels of the median centroid location. We then carry out the data reduction as described in \citet{Dittmann_1132}. After obtaining the pixel-level coefficients for our observations, we apply these coefficients to the unbinned and unnormalized data. We sum the values of these weighted pixels in order to obtain the total flux from LHS 1140. We apply a new outlier rejection criteria to these unbinned data. We reject all data points 7.5 median absolute deviations from a 12-minute wide sliding median in order to eliminate single point outliers, resulting in the rejection of 8 total data points (0.09\% of the total number of data points).

Our final data set consists of 135 sets of 64 individual sub-array images, each with an integration time of 2 seconds (our final datacube does not contain a full set of 64 images, as it was the last datacube of the observing program), for a total of 8545 data points. These data were calibrated with the {\emph{Spitzer}} pipeline version S19.2.0 and the timestamps of each data point are calculated at the Solar System barycenter in the TDB time system \citep{Eastman2010}.

\subsection{HARPS}

We intensively monitored LHS~1140 with the High Accuracy Radial velocity Planet Searcher (HARPS) in the frame of the \textit{M~dwarf program} (ESO Id: 191.C-0873(A), 198.C-0838(A); PI: Bonfils) and a survey dedicated to LHS~1140 (ESO Id: 0100.C-0884(A); PI: Astudillo-Defru). HARPS is a fiber-fed echelle spectrograph located at the 3.6m telescope at the La Silla observatory in Chile. Its resolving power is R=115,000, and it has a wavelength coverage between 380 nm and 690 nm over 72 orders \citep{Mayor2003}. HARPS achieves a sub-ms$^{-1}$ long-term precision thanks to its temperature and pressure controlled environment and a simultaneous wavelength calibration through a second fiber. We opted to place the calibration fiber on the sky to avoid contamination of the science spectra from a calibration spectrum, notably in the region of the Ca II H\&K lines. This choice does not degrade our RV precision, because for a star with the brightness of LHS 1140, our RV precision is dominated by the photon-noise of the stellar spectrum, as opposed to the calibration of the spectrograph.

We obtained 294 spectra of LHS~1140 between November 23, 2015 and January 15, 2018, spanning 784 days. In general, HARPS observations consist of two spectra per night with an exposure time of 1,800 seconds each. On some nights, we observed LHS~1140 one or three times. We discarded one observation (BJD: 2457686.6227) with an exposure time of only 1.8 sec, so our analysis was done over 293 radial velocity measurements.

The HARPS Data Reduction Software \citep[DRS,][]{Lovis2007} automatically reduces the spectra and computes the stellar radial velocity. The latter is done by cross-correlating spectra with a mask designed to match the spectral lines. The DRS uses a mask containing the vast majority of the absorption lines present in the spectrum of an M~dwarf, however, a non-negligible amount of lines are left out. To recover as much Doppler signal as possible, we follow the recipe described in \citet[][and references therein]{Astudillo-Defru2015}: the radial velocities derived by the DRS are used to shift all spectra to a common reference frame, then the median is computed to obtain an enhanced stellar spectrum (template). This template is Doppler shifted in a range of velocities to maximize its likelihood with each spectrum, resulting in the set of radial velocities used hereafter.

\section{Discovery of LHS 1140 \lowercase{c}} \label{sec:discovery}

\subsection{Machine learning to identify transit candidates} \label{sec:ann}

Articles describing the use of artificial neural networks to classify ground-based data in the exoplanet field have only recently appeared in the literature (known examples include \citet{Dittmann2017} and \citet{Armstrong2018}). Nevertheless, given that the first planet in the LHS 1140 system was discovered by \citet{Dittmann2017} via a machine learning process, we decided to use a similar approach to search for additional potential transit events in the MEarth data. MEarth generates many "triggers" per night (see Section \ref{sec:mearth}), most of which can be explained by variations in the atmosphere (clouds, changes in the column density of precipitable water vapour), rapid stellar variability (e.g. stellar flares), or instrumental systematics. Such false signals can often be identified by correlating the change in flux with a series of atmospheric and instrumental parameters such as the common mode (defined in Section \ref{sec:mearth}), the FWHM or pixel position of the target, or simultaneous variations in the fluxes of background stars. A small minority of triggers correspond to genuine transit events, which are not expected to be significantly correlated with any of these parameters. Our machine learning approach seeks to take advantage of this distinction - that is, we would like to eliminate as many triggers as possible based on systematics, and keep the rest as potential transit candidates.

After testing several neural network designs including feedforward and recurrent neural networks (see \citet{Murphy2012} for more information about neural networks), we settled for a simple feedforward neural network (FNN) consisting of an input layer, a single hidden layer, and an output node. The network design employed in this paper is similar to the one used by \citet{Dittmann2017} and can be summarized as follows. The input layer was composed of 10 neurons for atmospheric and instrumental parameters including the airmass, angle of the frame relative to the reference frame, RMS of the fluxes of background stars, and the FWHM, ellipticity and pixel position of the target, as well as the common mode with its time derivative and offset from a 3-hour median. The hidden layer was made up of 10 neurons as well, each fully connected to every input node. Finally, the FNN had a single output node representing the probability that the given input corresponds to a trigger. The data set, which was randomly split up into a training set (90\%) and a validation set (10\%), consisted of 1539 real triggers from MEarth between the years of 2014 and 2017. We complemented this set with an equal number of non-triggers which were selected randomly from the regular, low-cadence observations while ensuring that each star contributed the same amount of triggers and non-triggers (however, the number of triggers contributed by each star varied considerably). After training the network, we reclassified the 1539 real triggers, and selected the events where the FNN generated the lowest probabilities of being an expected trigger (that is, caused by atmospheric or instrumental effects) for further inspection. These events, unlike the majority of triggers, did not exhibit a detectable correlation with the atmospheric and instrumental parameters that we used as inputs.

\begin{figure}
	\includegraphics[width=\textwidth]{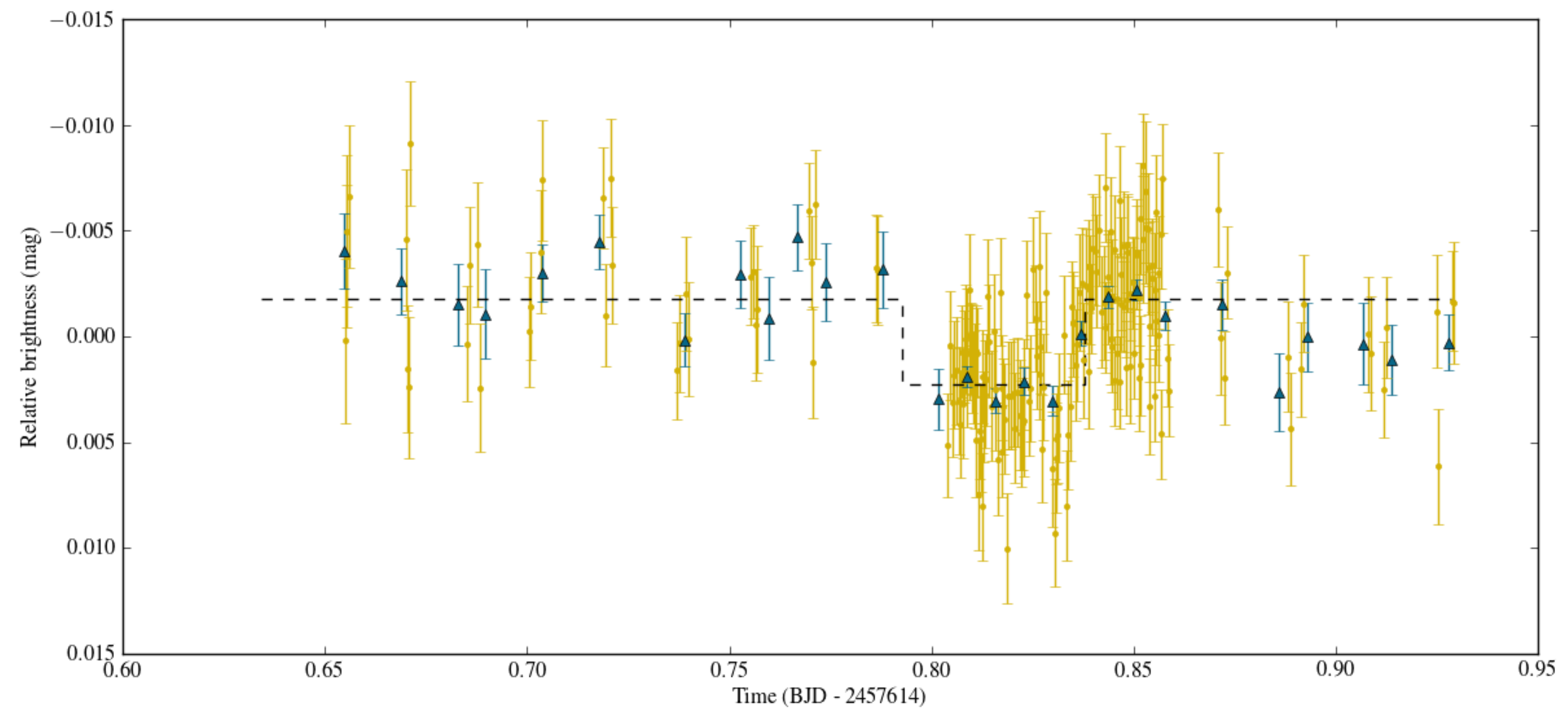}
	\caption{MEarth trigger event from August 14, 2016. The golden points represent individual measurements whereas the green triangles are weighted averages in 10-minute time bins. The dashed line illustrates the predicted transit of LHS 1140 c based on the results from joint photometric and RV modeling. According to the neural network, this event had a low probability of generating a trigger, meaning that the change in flux could not be explained by variations in the atmospheric and systematic parameters.\label{fig:trigger}}
\end{figure}

In particular, we identified two triggers belonging to LHS 1140 with very low trigger probabilities: one from 15 September 2014 UTC which was the basis of the LHS 1140 b discovery by \citet{Dittmann2017}, and another trigger from 14 August 2016 UTC (illustrated in Figure \ref{fig:trigger}) which did not overlap with any of the predicted transits of planet b. The estimated probabilities for the two events to be triggers were 23.2\% and 24.5\%, respectively, which placed them among the top 5\% events with the lowest trigger probabilities. The latter event was then selected as a potential transit candidate.

\subsection{BLS search} \label{sec:bls}

We also decided to carry out a phase-folded box model search in the photometry using the Box-fitting Least Squares (BLS) algorithm \citep{Kovacs2002}. In the case of LHS 1140, we tested over 1.6 million evenly spaced orbital frequencies corresponding to periods between 0.2 and 10 days. Our frequency spacing ensured that the maximum shift in orbital phase between two consecutive tested frequencies was always less than one half of the spacing between the time bins (we binned the observed fluxes into 10-minute error-weighted intervals). After fitting a box model to each orbital frequency, we imposed 4 selection criteria on the BLS spectrum to eliminate false detections. Firstly, we ignored any periods close to half a day or any integer number of days to avoid detecting variability that can be attributed to the Earth-based observing strategy. Secondly, we ignored any anti-transits (where the fitted transit depth was negative) as well as transit depths less than 0.1\%. We also required that the significance of the fitted model at a given period be at least 90\%, estimated by repeatedly refitting the data after adding normally distributed random offsets to the measured fluxes according to their respective uncertainties, and rejecting any outcomes that resulted in anti-transits. Finally, we calculated $\Delta \chi^2$, the difference in $\chi^2$ between the best-fit box model and a constant flux model, for each orbital frequency. In order to eliminate signals that are dominated by observations from a single night, we required that no night contribute more than 70\% of the total $\Delta\chi^2$-difference \citep{Burke2006}. We then ordered the remaining orbital frequencies in decreasing order of $\Delta\chi^2$.

\begin{figure}
	\includegraphics[width=\textwidth]{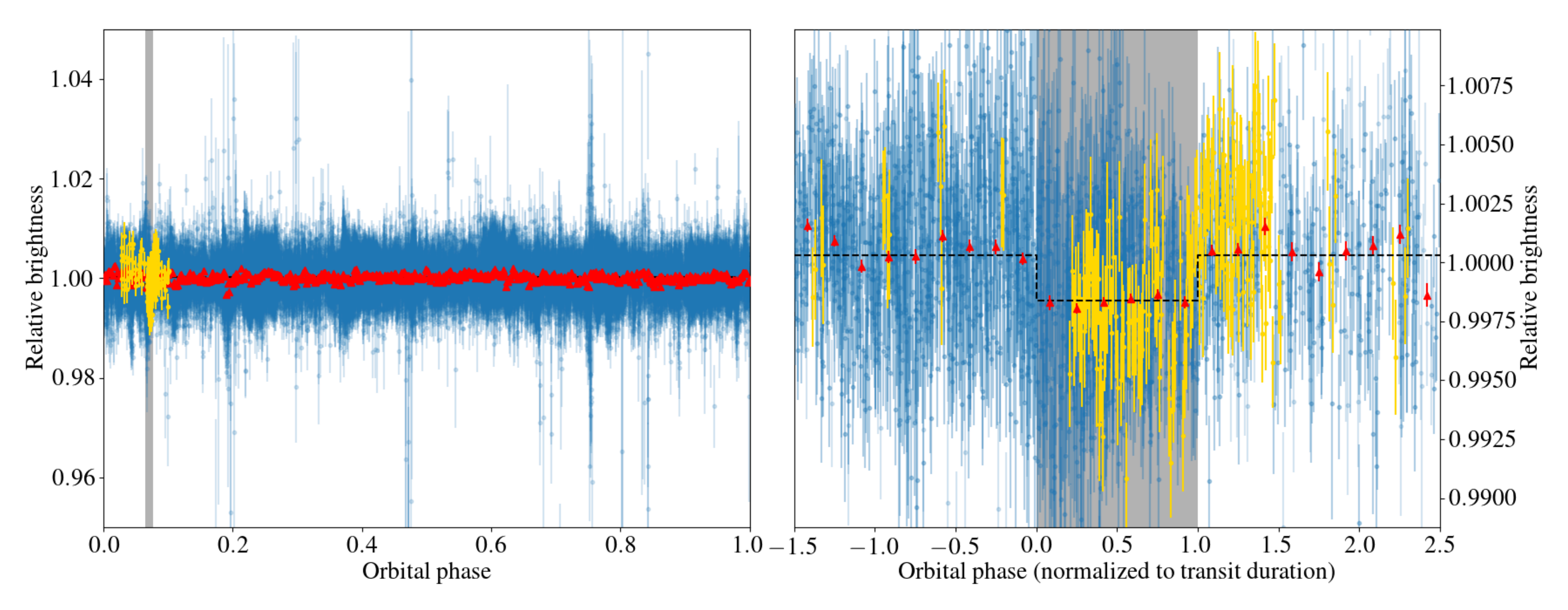}
	\caption{\textbf{Left:} A phase-folded display of the MEarth photometric data folded at a period of $P = 3.77797$ days. The yellow points represent the trigger from August 14, 2016, and the grey shaded area represents the proposed transits of LHS 1140 c. The red triangles are 10-minute binned averages. \textbf{Right:} A zoomed-in version of the left-side plot illustrating the transits of LHS 1140 c. The dashed line illustrates a fitted box model for the transits. The horizontal axis is rescaled to count the number of transit durations from the beginning of the transit.\label{fig:bls}}
\end{figure}

After masking out the predicted transits of LHS 1140 b, we investigated the 16 highest $\Delta\chi^2$-peaks in the BLS spectrum. One of these orbital periods, $P = 3.77797$ days, produced a phase-folded model that back-predicted a transit on August 14, 2016, with the ingress and egress times consistent with the MEarth trigger described in Section \ref{sec:ann}. Overall, the in-transit observations spanned 35 different nights, with no single night contributing more than 19\% to the total $\Delta\chi^2$-difference. The predicted transit depth from the box model was 0.192\% which corresponds to an Earth-sized planet. The phase-folded model is displayed in Figure \ref{fig:bls}.

\subsection{A transit of LHS 1140 c in the Spitzer data}

\begin{figure}
	\plotone{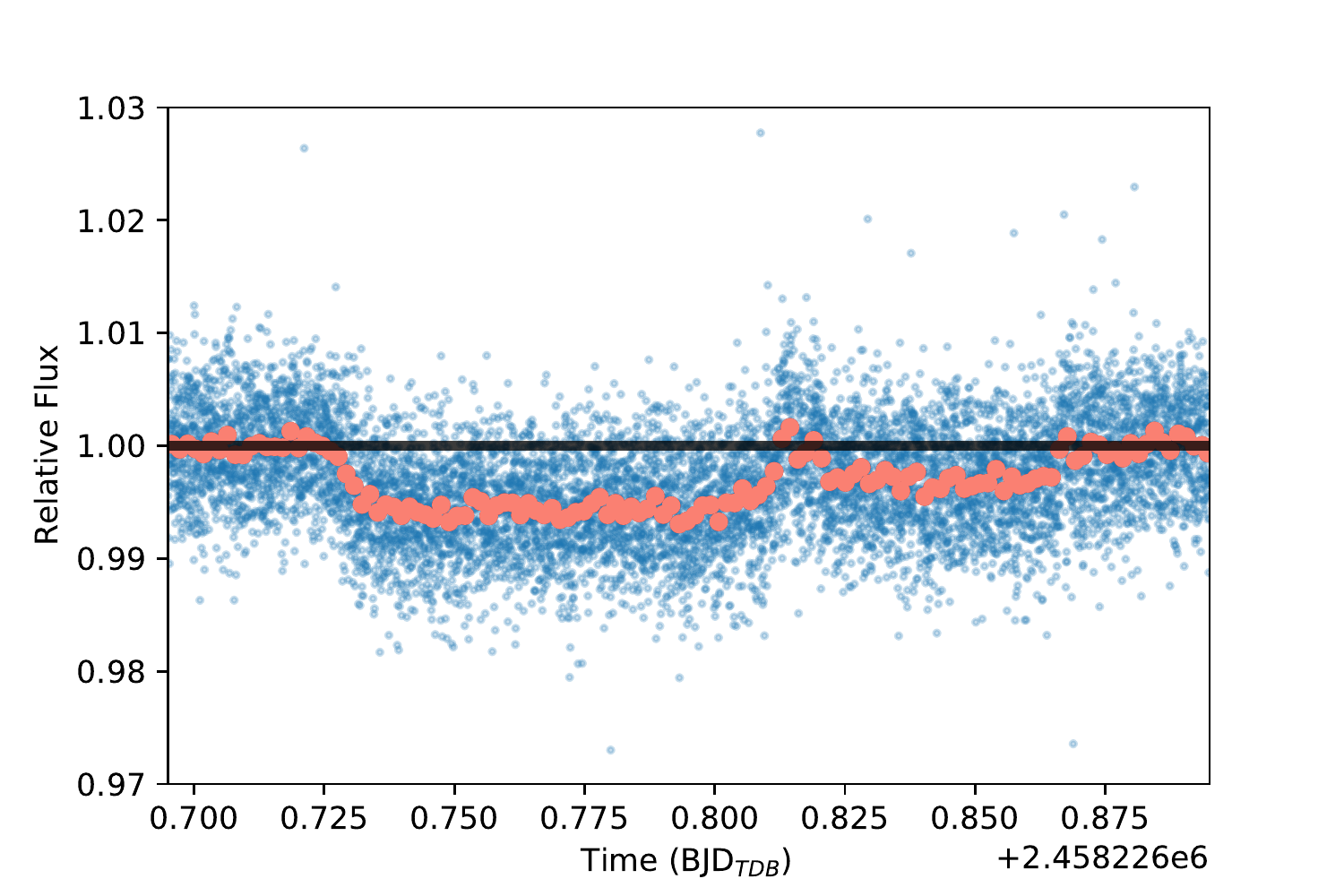}
	\caption{\emph{Spitzer} data of a transit of LHS 1140 b taken on April 18, 2018. Unbinned data are shown in blue and a binned version of the light curve is shown in salmon. The transit of LHS 1140 b is visible as the first dip in the plot. Following the egress of LHS 1140 b, we serendipitously observed a transit of LHS 1140 c, visible as the second smaller and shorter transit in the light curve. Members of our team recovered this planet independently in the \emph{Spitzer} and MEarth photometry as well as in the HARPS radial velocity measurements.}
	\label{fig:Spitzer_lightcurve_with_the_sweet_color_scheme}
\end{figure}

We show our calibrated \emph{Spitzer} data in Figure \ref{fig:Spitzer_lightcurve_with_the_sweet_color_scheme}. In this plot, we see the transit of LHS 1140b followed by a second transit event of smaller depth and shorter duration. This second transit event is not associated with shifts in the pixel position of the centroid of the target. The duration of this event is approximately 74 minutes with a depth of 3 mmag. 

The discovery of the \emph{Spitzer} transit event (by J.A.D.) happened independently of identifying the trigger or phase-folded transits of LHS 1140 c in the MEarth data (led by K.M.) and independently of the HARPS radial velocity identification of LHS 1140 c (led by N.A-D.). We discuss the simultaneous fitting of these transits with radial velocity observations in Section \ref{sec:jointfit}. 

\subsection{Periodogram analysis of the RVs}

\begin{figure}
	\includegraphics[width=\textwidth]{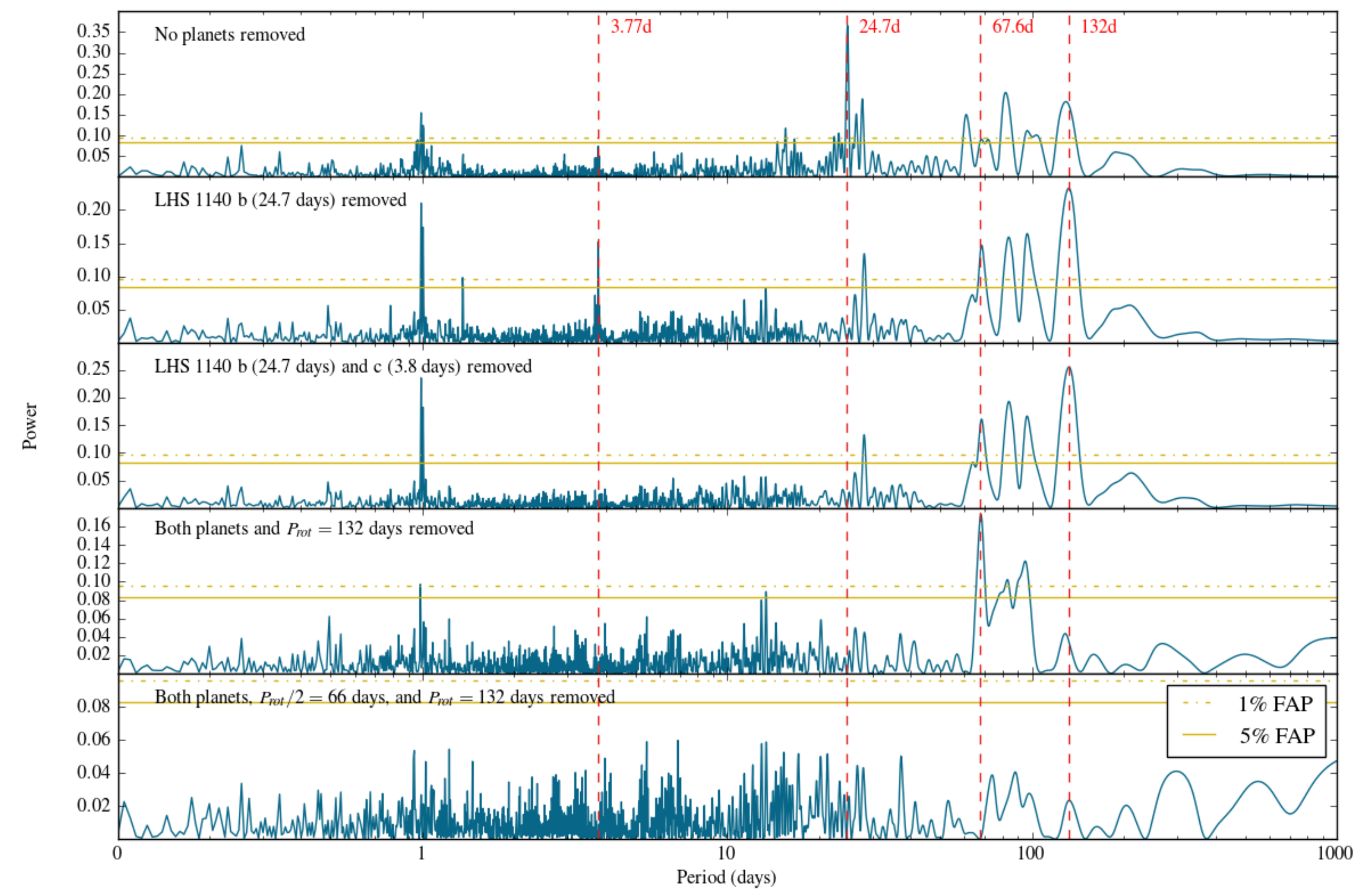}
	\caption{A Generalized Lomb-Scargle (GLS) periodogram of the radial velocities. The top panel shows the GLS for the original RV data set, whereas the second and the third panels illustrate the GLS of the residuals after removing one or both planets, respectively. The residuals of the two-planet model exhibit a series of significant GLS peaks at periods of 65 days and higher. However, they can be removed by introducing a model for stellar activity which consists of the first two harmonics of the stellar rotation period, as shown in the two bottom panels. As a result, we associate these peaks with stellar activity. Vertical dashed lines are overplotted on all panels to represent the periods of every modeled signal (both planetary and activity-induced). Horizontal lines estimate the 1\% and 5\% false alarm probabilities (FAP) to eliminate any peaks arising from window functions.\label{fig:periodogram}}
\end{figure}

Using HARPS radial velocities, we unveiled the signal of LHS~1140~c independently of the photometry. A Generalized Lomb-Scargle \citep[GLS,][]{Zechmeister2009} periodogram analysis of the RVs reveals several interesting signals, shown in Figure \ref{fig:periodogram}. The most significant peak lies close to 25 days which corresponds to the orbital period of LHS 1140 b. After subtracting a Keplerian model for LHS 1140 b from the RVs, a strong signal emerges near the stellar rotation period $P_{rot} = 131 \pm 5$ days as well as the 3.8-day period, matching the one found by the BLS analysis in Section \ref{sec:bls}. After subtracting a two-planet model, we produced a periodogram displayed in the third panel of Figure \ref{fig:periodogram}, exhibiting several low-frequency peaks at around 65 days and longer periods. At least two of these peaks, at 68 days and 132 days, can be attributed to the stellar rotation period \prot\ and its harmonic \prot/2. Collections of starspots can often be successfully modeled as a series of no more than the first three or four harmonics of \prot\ \citep{Jeffers2009,Boisse2011,Haywood2015}, such as in the cases of Corot-7 \citep{Queloz2009,Boisse2011} and HD 189733 \citep{Boisse2011}. Therefore, we proceeded to generate a model for stellar activity that consists of a sum of two sinusoidal signals with periods initialized at \prot\ and \prot/2 (adding in further harmonics was not necessary in this case). Subtracting this model from the residuals significantly reduces any residual power in the low-frequency signals beyond 65 days (displayed in the bottom panel of Figure \ref{fig:periodogram}). Thus, we see no evidence that the previously significant GLS peaks between 65 and 130 days were caused by anything other than stellar activity.

\section{Parameter estimation for LHS 1140 \lowercase{b} and \lowercase{c}} \label{sec:modeling}

\begin{deluxetable}{lccc}
    \tabletypesize{\footnotesize}
	\tablewidth{0pt}
	\tablecaption{System parameters for LHS 1140\label{tbl:results}}
	\tablehead{
		Parameter & \multicolumn{2}{c}{Values for LHS 1140} & Source\tablenotemark{a}
	}
	\startdata
	\multicolumn{3}{l}{\textbf{Stellar parameters}}\\
	Right ascension (J2000) & \multicolumn{2}{c}{00h 44min 59.3s} & (1)\\
	Declination (J2000) & \multicolumn{2}{c}{-15$^\circ$ 16' 18"} & (1)\\
	Proper motion (mas yr$^{-1}$) & \multicolumn{2}{c}{
	    $\mu_\alpha = 317.59 \pm 0.12 \quad \mu_\delta = -596.62 \pm 0.09$} & (1)\\
	Apparent brightness (mag) & \multicolumn{2}{c}{
	    \begin{tabular}{cc}
	        $V_J = 14.18 \pm 0.03$ & $J = 9.612 \pm 0.023$\\
	        $R_C = 12.88 \pm 0.02$ & $H = 9.092 \pm 0.026$\\
	        $I_C = 11.19 \pm 0.02$ & $K_S = 8.821 \pm 0.024$
	    \end{tabular}
	} & (2)(4)\\
	Distance (pc) & \multicolumn{2}{c}{14.993 $\pm$ 0.015} & (1)\\
	Mass (\msun) & \multicolumn{2}{c}{0.179 $\pm$ 0.014} & (3)\\
	Radius (\rsun) & \multicolumn{2}{c}{0.2139 $\pm$ 0.0041} & (3)\\
	Luminosity (\lsun) & \multicolumn{2}{c}{0.00441 $\pm$ 0.00013} & (3)\\
	Effective temperature (K) & \multicolumn{2}{c}{3216 $\pm$ 39} & (3)\\
	Metallicity [Fe/H] & \multicolumn{2}{c}{-0.24 $\pm$ 0.10} & (4)\\
	Age (Gyr) & \multicolumn{2}{c}{$>5$} & (4)\\
	Rotational period (days) & \multicolumn{2}{c}{131 $\pm$ 5} & (4)\\
	\hline\hline
	Parameter & LHS 1140 b & LHS 1140 c\\
	\hline
	\multicolumn{3}{l}{\textbf{Modeled transit and RV parameters}}\\
	Orbital period $P$ (days) & 24.736959 $\pm$ 0.000080 & 3.777931 $\pm$ 0.000003\\
	RV semi-amplitude $K$ (\ms) & 4.85 $\pm$ 0.55 & 2.35 $\pm$ 0.49\\
	Eccentricity $e$ (90\% confidence) & $<0.06$ & $<0.31$\\
	Time of mid-transit $t_{\rm T}$ (BJD) & 2456915.71154 $\pm$ 0.00004 & 2458226.843169 $\pm$ 0.000026\\
	Inclination $i$ (deg) & $89.89^{+0.05}_{-0.03}$ & $89.92^{+0.06}_{-0.09}$\\
	Planet-to-star radius ratio $r/R_\star$ & 0.07390 $\pm$ 0.00008 & 0.05486 $\pm$ 0.00013\\
	$a/R_\star$ ratio & 95.34 $\pm$ 1.06 & 26.57 $\pm$ 0.05\\
	\hline
	\multicolumn{3}{l}{\textbf{Derived planetary parameters}}\\
	Mass $m$ (\mearth) & 6.98 $\pm$ 0.89 & 1.81 $\pm$ 0.39\\
	Radius $r$ (\rearth) & 1.727 $\pm$ 0.032 & 1.282 $\pm$ 0.024\\
	Density $\rho$ (g cm$^{-3}$) & 7.5 $\pm$ 1.0 & 4.7 $\pm$ 1.1\\
	Surface gravity $g$ (m s$^{-2}$) & 23.7 $\pm$ 2.7 & 10.6 $\pm$ 2.2\\
	Semi-major axis $a$ (AU) & 0.0936 $\pm$ 0.0024 & 0.02675 $\pm$ 0.00070\\
	Incident flux $S$ (\searth) & 0.503 $\pm$ 0.030 & 6.16 $\pm$ 0.37\\
	Equilibrium temperature\tablenotemark{b} $T_{\rm eq}$ (K) & 235 $\pm$ 5 & 438 $\pm$ 9\\
	\enddata
	\tablenotetext{a}{(1) \citet{Gaia2018}, (2) \citet{2MASS}, (3) This work, (4) \citet{Dittmann2017}.}
	\tablenotetext{b}{The equilibrium temperature assumes a Bond albedo of 0. For an albedo of $A_{B}$, the reported temperature needs to be multiplied by $(1-A_{B})^{1/4}$.}
\end{deluxetable}

\subsection{A joint photometry and RV analysis} \label{sec:jointfit}

We fit a joint model of the \emph{Spitzer} observations, MEarth-South observations, and the radial velocity measurements from HARPS to simultaneously constrain the masses, radii, and orbital parameters of both planets, as well as the radial velocity variability due to the stellar activity of LHS 1140. Our \emph{Spitzer} data are shown in Figure \ref{fig:Spitzer_lightcurve_with_the_sweet_color_scheme} and described in Section \ref{sec:spitzer}. We have observed one transit of LHS 1140 c with \emph{Spitzer} and we present one transit of LHS 1140 b with \emph{Spitzer} here as well. The remaining transits of LHS 1140 b from the \emph{Spitzer} program will be the subject of a future publication. We do not currently have a full transit observation of LHS 1140 c with MEarth-South, but we fit the transit observations obtained on 35 individual transit nights (see section \ref{sec:bls}), including the night of the trigger.

We fit our photometric observations with the \texttt{batman} code \citep{batman}. \texttt{batman} is an optimized python implementation of the analytical model for transit light curves from \citet{mandel_and_agol} and its erratum. We model the radial velocity variations of LHS 1140 from the orbits of the two planets (consistent with the transit ephemeris) and we adopt a simple sinusoidal model to describe the radial velocity variations of LHS 1140 due to the activity on the star. 

We initiate the physical parameters for LHS 1140 b with the values found in \citet{Dittmann2017}. We adopt limb darkening coefficients from \citet{claret2012_limb_darkening} for a 3300 K star with log(g)$ = 5.0$ in a Cousins $I$ filter. The Cousins $I$ filter has a similar effective wavelength to the MEarth bandpass. For the \emph{Spitzer} transits, we adopt limb darkening coefficients from \citet{Spitzer_Claret}, which are calculated for the \emph{Spitzer} bandpass. We initiate the period of LHS 1140 c at the best fit period in the phase-folded detection and a $T_0$ estimated from the midpoint of the \emph{Spitzer} transit. The radial velocity amplitude for LHS 1140 c is initialized to an amplitude of 3 m s$^{-1}$, which is the approximate amplitude a rocky composition planet would exhibit at this orbital period. 
In order to efficiently explore parameter space, we use the \texttt{emcee} code \citep{emcee}. \texttt{emcee} is a python implementation of the Affine-Invariant Markov Chain Monte Carlo sampler \citep{GoodmanWeare2010}. Each model is initiated with 100 walkers in a Gaussian ball located at this initial solution. We run each chain for 100,000 steps and discard the first 10\% of steps so that the solution may ``burn-in" independent of the choice of initialization. We show the results of this analysis in Table \ref{tbl:joint_fit}.

\begin{deluxetable}{lll}
	\tablewidth{0pt}
	\tablecaption{Model parameters from the joint photometry and radial velocity fit in Section \ref{sec:jointfit} \label{tbl:joint_fit}}
	\tablehead{
		\colhead{Parameter} & \colhead{Value} & \colhead{Explanation}
	}
	\startdata
	$P_b$ (days) & $24.736959 \pm 0.000080$ & Orbital period of LHS 1140 b \\
	$T_{0,b}$ (BJD) & $2456915.71154 \pm 0.000040$ & Transit time of LHS 1140 b \\
	$\frac{r_b}{r_*}$ & $0.07390 \pm 0.00008$ & Ratio of LHS 1140 b and LHS 1140's radii \\
	$\frac{a_b}{r_*}$ & $95.34 \pm 1.06$ & Ratio of semi-major axis to stellar radius for LHS 1140 b\\
	$i_b$ (degrees) & $89.89^{+0.05}_{-0.03}$ & Inclination angle of LHS 1140 b\\
	$e_b$ & 0 (fixed; $<0.094$ at 99\% confidence) & Eccentricity of LHS 1140 b\tablenotemark{a}\\
	$K_b$ (m s$^{-1}$) & $4.71 \pm 0.09$ & RV amplitude of LHS 1140 b\tablenotemark{a}\\
	$P_c$ (days) & $3.777931^{+0.000004}_{-0.000002}$ & Orbital period of LHS 1140 c \\
	$T_{0,c}$ (BJD) & $2458226.843169 \pm 0.000026$ & Transit time of LHS 1140 c \\
	$\frac{r_c}{r_*}$ & $0.05486 \pm 0.00013$ & Ratio of LHS 1140 c and LHS 1140's radii \\
	$\frac{a_c}{r_*}$ & $26.57 \pm 0.05$ & Ratio of semi-major axis to stellar radius for LHS 1140 c \\
	$i_c$ (degrees) & $89.92^{+0.06}_{-0.09}$ & Inclination angle of LHS 1140 c \\
	$e_c$ & 0 (fixed; $<0.236$ at 99\% confidence) & Eccentricity of LHS 1140 c\tablenotemark{a}\\
	$K_c$ (m s$^{-1}$) & $2.42 \pm 0.10$ & RV amplitude of LHS 1140 c\tablenotemark{a}\\
	$\gamma$ (km s$^{-1}$) & $-13.23851 \pm 0.00008$ & RV zeropoint of LHS 1140\tablenotemark{a}\\
	$K_*$ (m s$^{-1}$) & $3.0 \pm 0.1$ & RV amplitude of stellar activity\tablenotemark{a}\\
	$\phi_*$ (BJD) & $2457766.66 \pm 0.68$ & RV phase of star\tablenotemark{a}\\
	$P_*$ (days) & $132.26 \pm 0.37$ & RV period of star\tablenotemark{a}\\
	$a_{4.5}$ & $0.0104 \pm 0.0007$ & \emph{Spitzer} limb darkening coefficient \\
	$b_{4.5}$ & $0.2175^{+0.0012}_{-0.0042}$ & \emph{Spitzer} limb darkening coefficient \\
	$a_{\rm MEarth}$ & $0.195^{+0.022}_{-0.007}$ & MEarth limb darkening coefficient \\
	$b_{\rm MEarth}$ & $0.356^{+0.013}_{-0.11}$ & MEarth limb darkening coefficient
	\enddata
	\tablenotetext{a}{These values were merely used as intermediate results, and were subsequently supplanted by the values from the Gaussian Process modeling, listed in Table \ref{tbl:results}.}
\end{deluxetable}

\begin{figure}
	\plottwo{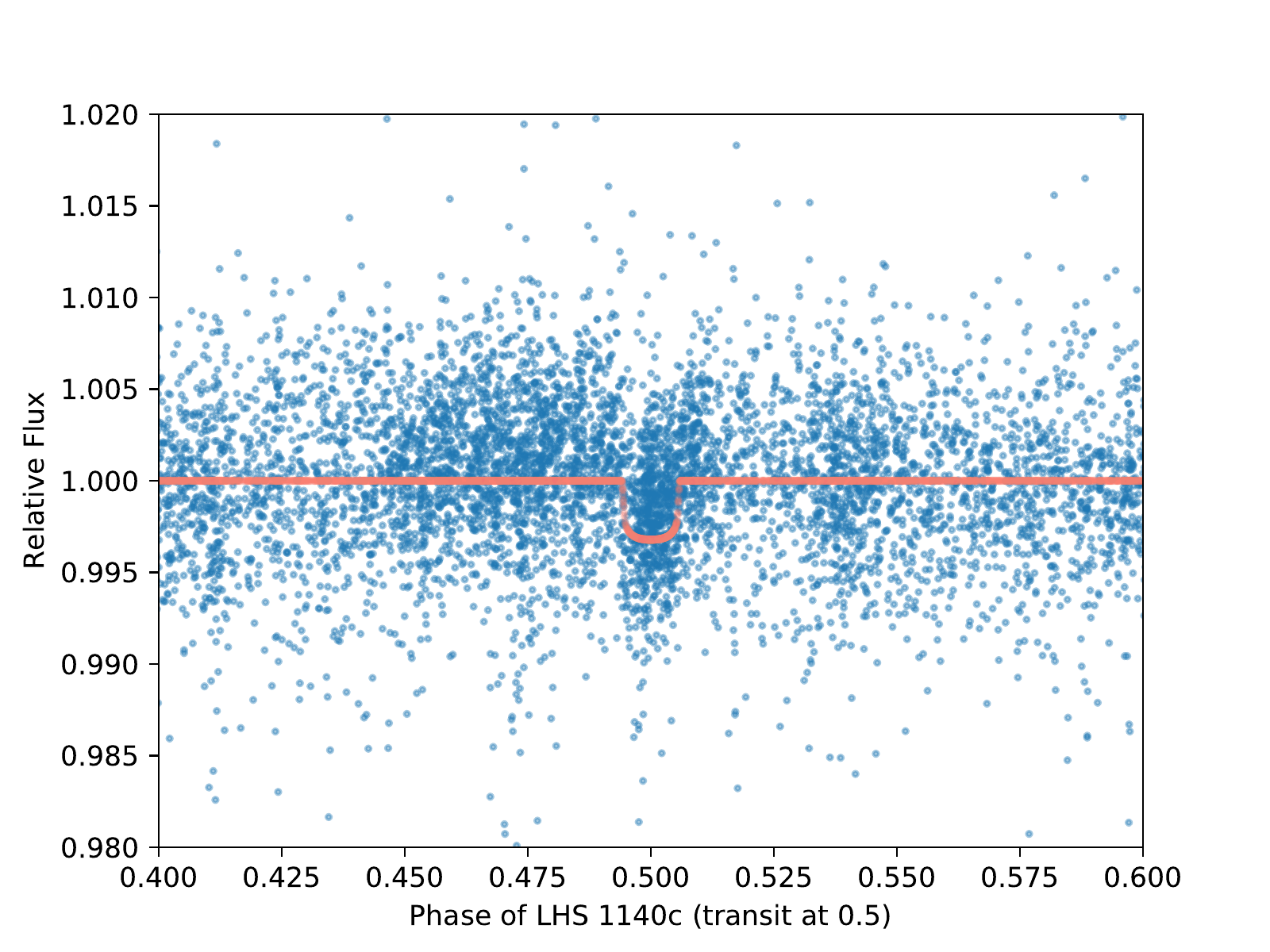}{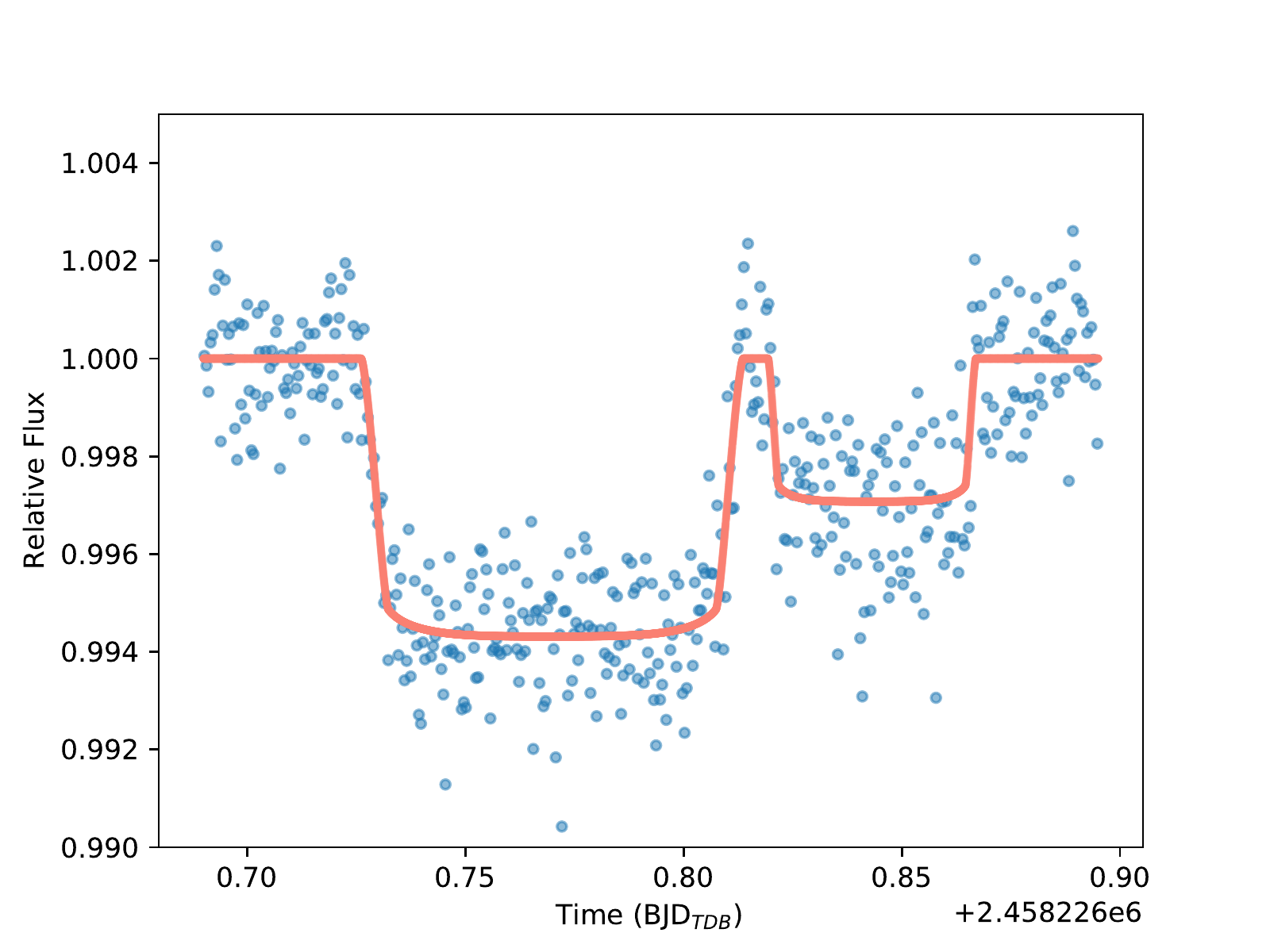}
	\caption{\textbf{Left}: MEarth-South photometry of LHS 1140 phased to the orbit of LHS 1140 c. The salmon line is our joint photometry and radial velocity fit. Here we have removed all transits of LHS 1140 b. \textbf{Right}: Binned \emph{Spitzer} photometry of the double transit of LHS 1140 b (left dip) and LHS 1140 c (right dip). The salmon line is our joint model fit for the system.}
	\label{fig:joint_photometry}
\end{figure}

\begin{figure}
	\plottwo{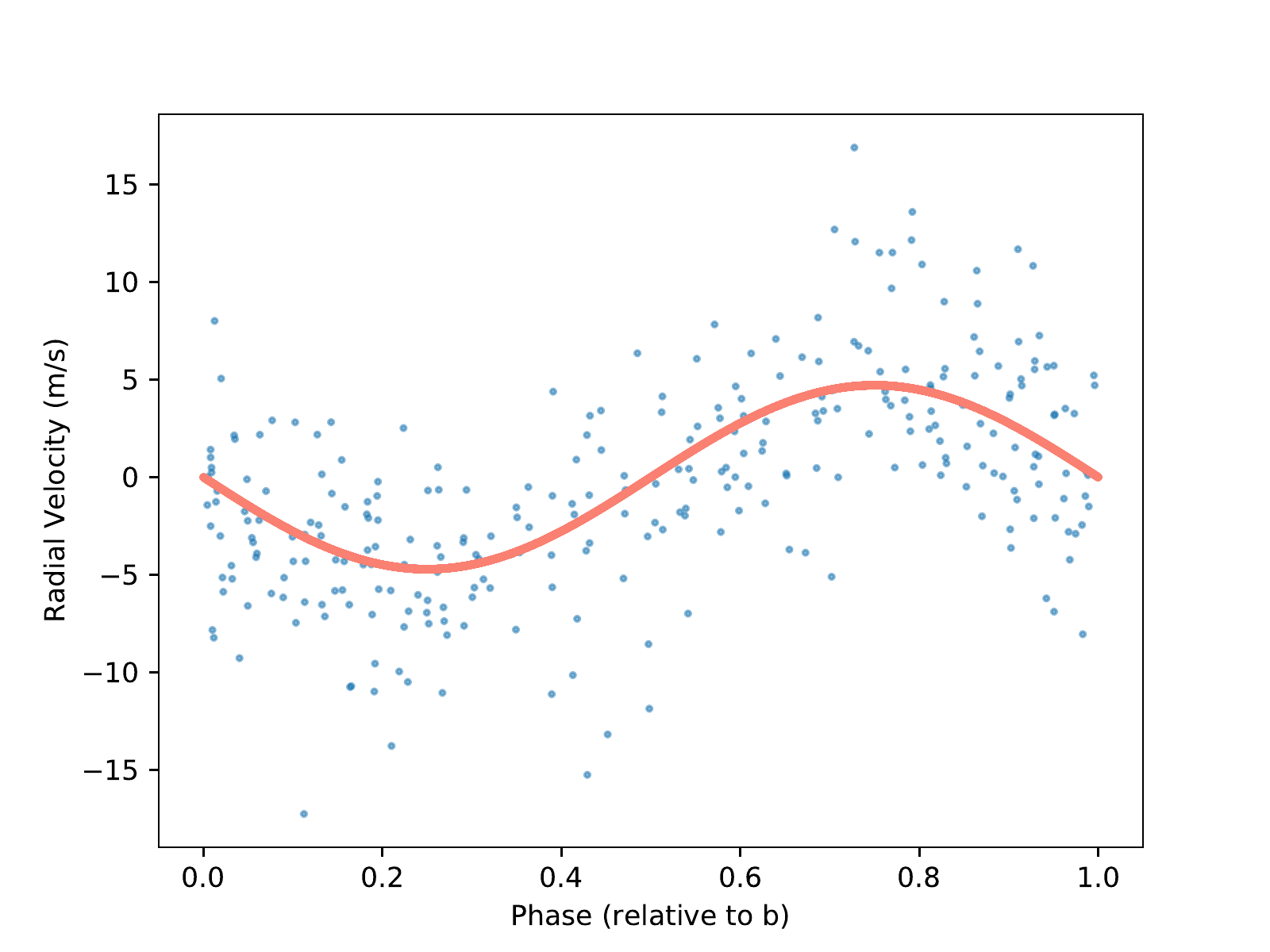}{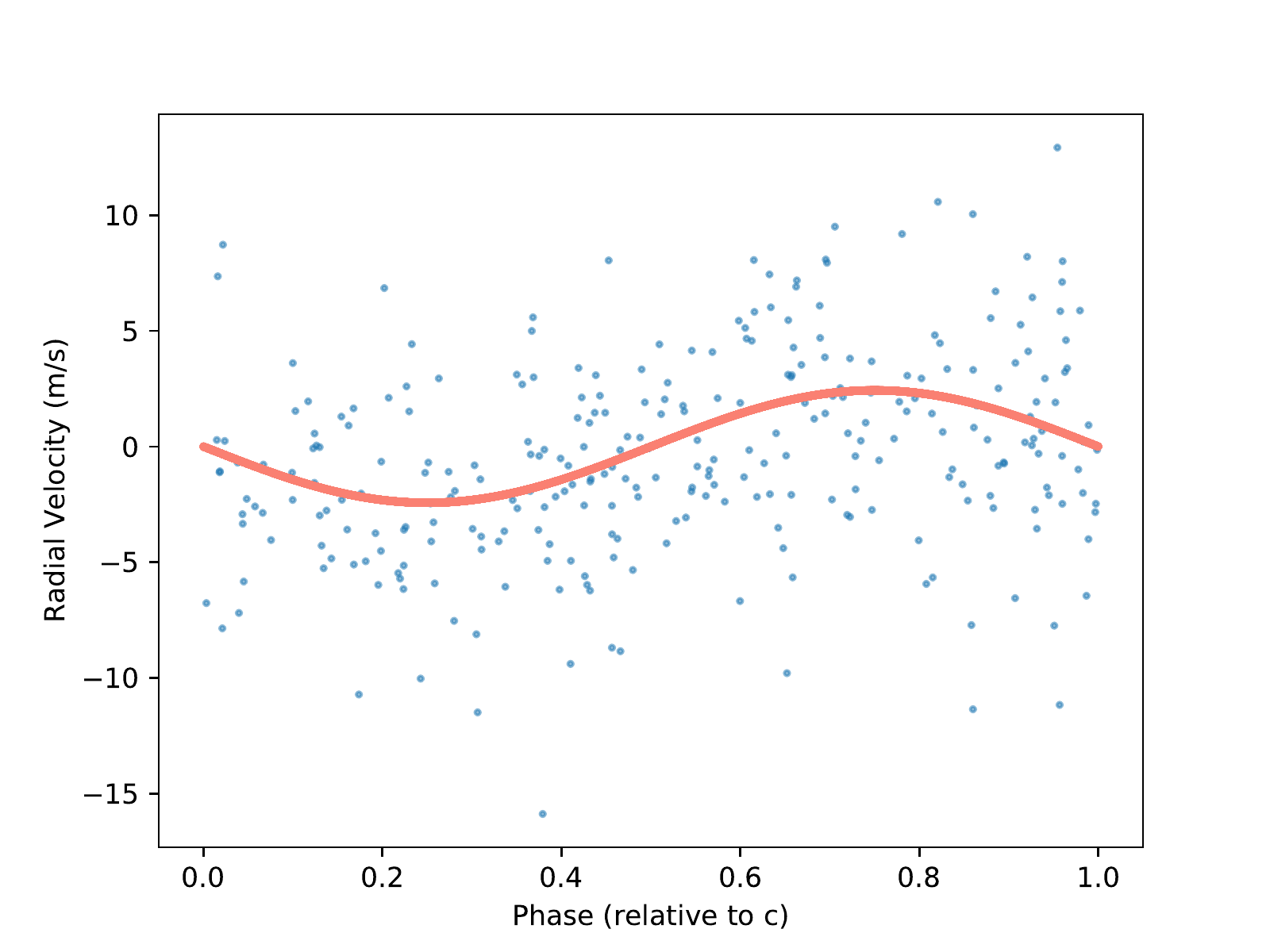}
	\caption{Radial velocity measurements phased to the orbit of LHS 1140 b (\textbf{left}) and LHS 1140 c (\textbf{right}). Data points are in blue with the model radial velocity curve shown in salmon. For each curve we have removed the effects of stellar radial velocity and the radial velocity induced by the other planet. We find no evidence for eccentricity in either planet.}
	\label{fig:joint_rv}
\end{figure}

We show the results of these fits in Figures \ref{fig:joint_photometry} and \ref{fig:joint_rv}. We are able to recover the phased transits of LHS 1140 c in the MEarth-South photometry and simultaneously fit the photometry and radial velocities for both planets. Our planetary parameters for LHS 1140 b are consistent with, but more precise than, those presented in \citet{Dittmann2017}. We find that LHS 1140 c has a transit depth of 3.0 mmag. We present further discussion of the results of these fits and the physical parameters of the LHS 1140 system in Section \ref{sec:discussion}.

Up to this point, we have modeled the radial velocities using a simplistic model: we have only considered the effects of both of the planets in the system and have modeled the stellar signal as a single sinusoid at the stellar rotation period (measured through our photometry). This analysis assumes that each data point is independent and that there is no correlated noise on any timescale. We now relax these assumptions to more accurately capture the correlated noise and errors present in our radial velocity measurements.

\subsection{Remodeling the RVs with Gaussian process regression}
As demonstrated by the GLS periodogram in Figure \ref{fig:periodogram}, the radial velocities exhibit significant correlation near the stellar rotation period of $131 \pm 5$ days as well as its harmonics. The exact properties of these quasi-periodic variations in M dwarfs are yet to be resolved, although they are commonly believed to be linked to features on the stellar surface such as dark spots and granulation. Overall, a constant sinusoidal model (as presumed in Section \ref{sec:jointfit}) might be an inaccurate representation of activity-induced RV modulations. In pursuit of a more accurate accounting of the latter, we employed Gaussian process (GP) regression \citep{Rasmussen2006} which has recently been applied to several exoplanetary systems including Corot-7 \citep{Haywood2014,Faria2016}, Kepler-21 \citep{LopezMorales2016}, Kepler-1655 \citep{Haywood2018}, K-2 18 \citep{Cloutier2017}, Alpha Centauri B \citep{Rajpaul2015}, and LHS 1140 \citep{Dittmann2017}. Our goal in using GP regression is to make as few assumptions about the exact form of the signal due to stellar activity as possible. We did not include the photometric data in our GP framework in order to make the analysis computationally manageable. Our GP regression was implemented by using a quasi-periodic kernel of the form:
\begin{equation}
    k(t, t') = \eta_1^2 \cdot \exp \left[ -\frac{(t - t')^2}{\eta_2^2} - \frac{\sin^2 \left( \frac{\pi (t - t')}{\eta_3} \right) }{\eta_4^2} \right]
\end{equation}
where $\eta_1$ represents the amplitude of the covariance function, $\eta_2$ is the evolution timescale of activity-inducing stellar surface features, $\eta_3$ is the recurrence timescale (the expected period of correlation), and $\eta_4$ determines the amount of high-frequency structure in the GP. The Gaussian process framework is extremely versatile, and therefore it is often desirable to constrain the values of the hyperparameters with tight priors based on physical intuition about the system. We decided to adopt our priors from \citet{Dittmann2017} for consistency purposes. As a result, the recurrence timescale $\eta_3$ was constrained by a Gaussian prior centered at the stellar rotation period ($131 \pm 5$ days). The active regions of the star are likely to remain stable over a few rotation periods, so we adopt a Gaussian prior at three times the rotation period (393 $\pm$ 30 days) for the evolution timescale $\eta_2$. Similarly to \citet{Dittmann2017}, choosing a different value of the same order of magnitude does not significantly change our results. The structure parameter $\eta_4$ is constrained by a Gaussian prior centered at $0.5 \pm 0.05$ which allows for two or three activity-induced peaks in the RV curve per rotation \citep{LopezMorales2016,Haywood2018}. RV models with more than three peaks per rotation would likely be unphysical due to the limb darkening and foreshortening effects that effectively erase any structure in the photometric and RV signatures \citep{Jeffers2009}.

We represented each planet $i$ by a Keplerian model with an amplitude $K_i$, an orbital period $P_i$, and a time of transit $t_{T,i}$. The amplitudes $K_i$ (as well as $\eta_1$) were constrained only by a modified Jeffreys prior. In order to accommodate the results of the joint fit from Section \ref{sec:jointfit}, we implemented tight Gaussian priors on $P_i$ and $t_{T,i}$ centered at the joint fit results while setting the widths equal to the respective uncertainties. Initially, our models also included eccentricities $e_i$ and arguments of periapsis $\omega_i$. However, we found the improvement in likelihood statistically negligible: the Bayesian information criterion (BIC) increased from 1173 to 1218 when substituting a two-planet eccentric model for a circular one. Thus, we used the eccentric model only to place an upper limit on the orbital eccentricities: we optimized the rest of the parameters assuming $e = 0$. The parameters were optimized via an affine-invariant Markov Chain Monte Carlo (MCMC) routine \citep{GoodmanWeare2010}. We present a list of all modeled parameters with their appropriate prior distributions in Table \ref{tbl:priors}.

Our best fit yielded an amplitude of $K_b = 4.85 \pm 0.55$ \ms\ and a period of $P_b = 24.73712 \pm 0.00032$ days for planet b. Similarly for planet c, we obtain $K_c = 2.35 \pm 0.49$ \ms and $P_c = 3.777950 \pm 0.000007$ days. Using these results, we can estimate the planetary masses as $m_b = 6.98 \pm 0.89$ \mearth\ and $m_c = 1.81 \pm 0.39$ \mearth\ (including the propagated uncertainties from the stellar mass). We also obtained a GP amplitude of $\eta_1 = 2.68 \pm 0.81$ \ms, GP timescales of $\eta_2 = 400 \pm 38$ days and $\eta_3 = 134.8 \pm 3.2$ days, and a structure parameter value of $\eta_4 = 0.51 \pm 0.06$. The latter three are all consistent with our initial priors. Using the eccentric models, we can place upper limits of $e_b < 0.060$ and $e_c < 0.313$ on the orbital eccentricities with 90\% confidence. However, we note that the posterior probability distribution for the eccentricity of planet c is significantly lopsided, with the 80\% confidence limit at $e_c < 0.14$. We decided to retain the orbital periods and times of transit from the combined fit in Section \ref{sec:jointfit} since these are constrained much more tightly by photometry. The uncertainties on all of the parameters were obtained by calculating the 14th and 86th percentiles of the final posterior distributions, and the best-fit values were taken from the medians of the PDFs.

\begin{figure}
	\includegraphics[width=\textwidth]{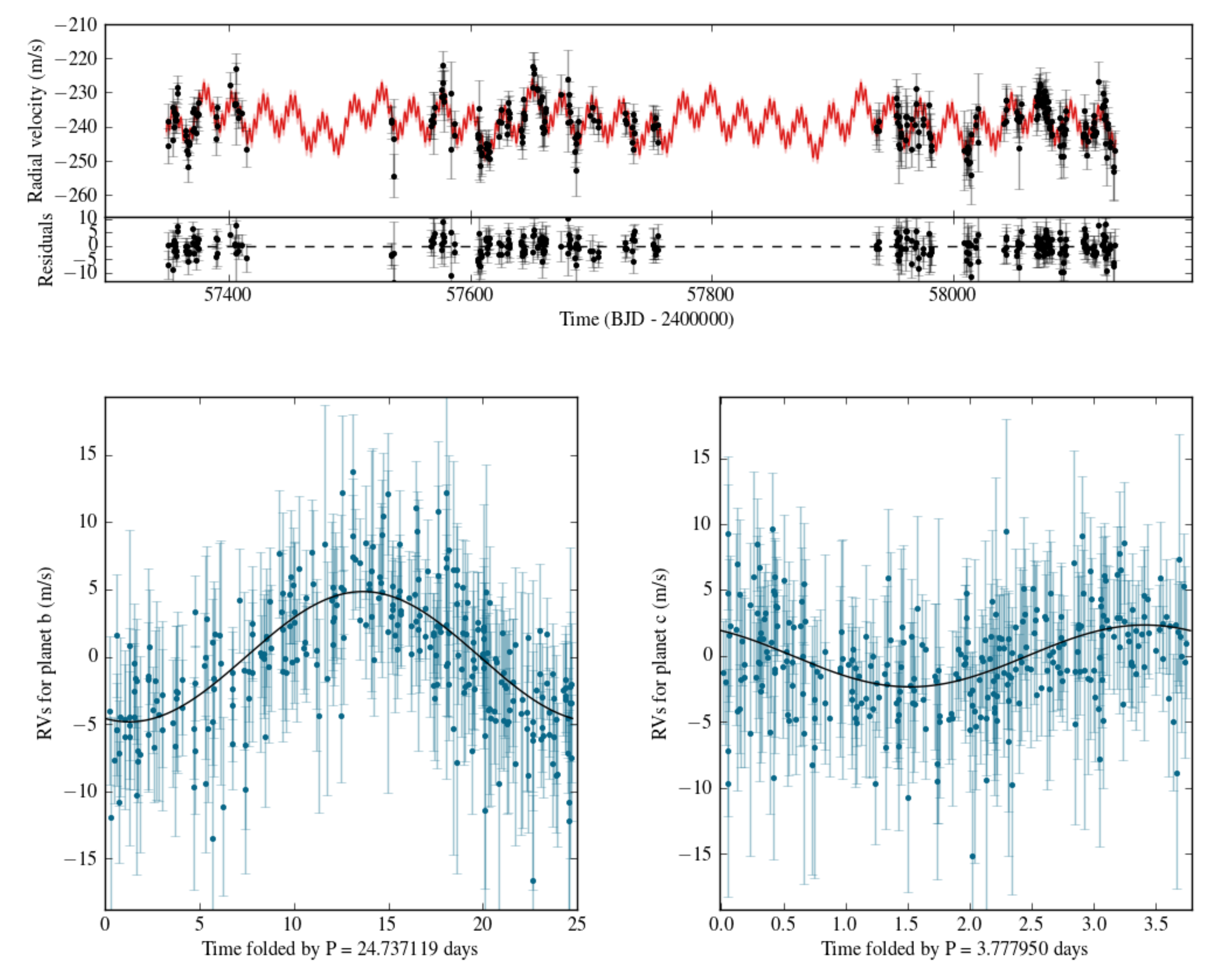}
	\caption{The best 2-planet model (with circular orbits) for the radial velocities. The top panels show the full model (and the corresponding RV residuals) including the two planets as well as an activity-induced offset from Gaussian Process (GP) regression. The bottom panels show the signals for each planet separately, with the other planet as well as the GP-predicted offset subtracted out, folded at the appropriate orbital periods.\label{fig:gp}}
\end{figure}

Our best circular RV model is illustrated in Figure \ref{fig:gp}, and our final values for all modeled and derived parameters are given in Table \ref{tbl:results}. We note that the 1$\sigma$ uncertainties for $K_b$ and $K_c$ from our updated RV solution are now consistent with the expected uncertainties based on the Fisher information. Previously, \citet{Cloutier2018} found that the measurement precision of $K_b$ in the 1-planet model of \citet{Dittmann2017} was about $\sqrt{2}$ times larger than its theoretical value based on the number of RV data points.

\begin{deluxetable}{ll}
	\tablewidth{0pt}
	\tablecaption{Modeled parameters and their prior distributions \label{tbl:priors}}
	\tablehead{
		\colhead{Parameter} & \colhead{Prior distribution}
	}
	\startdata
	$P$ (days) & Gaussian $(24.736959, 0.000080), (3.777931, 0.000003)$\\
	$K$ (\ms) & Modified Jeffreys $(\sigma_{\rm RV})$\\
	$e$ & Uniform $[0, 1)$\\
	$\omega$ (deg) & Uniform $[0, 360)$\\
	$t_{\rm T}$ (BJD) & Gaussian $(2456915.71154, 0.00004), (2458226.843169, 0.000026)$\\
	$\eta_1$ (\ms) & Modified Jeffreys $(\sigma_{\rm RV})$\\
	$\eta_2$ (days) & Gaussian (393, 30)\\
	$\eta_3$ (days) & Gaussian (131, 5)\\
	$\eta_4$ & Gaussian (0.5, 0.05)\\
	RV$_0$ & Uniform $(-\infty, +\infty)$\\
	\enddata
\end{deluxetable}

\section{Discussion and conclusion} \label{sec:discussion}

\citet{Dittmann2017} first announced the discovery of a transiting super-Earth orbiting the nearby M dwarf LHS 1140. We managed to better constrain the properties of the planet by estimating its mass $m_b = 6.98 \pm 0.89$ \mearth\ and radius $r_b = 1.727 \pm 0.032$ \rearth. These improved values are significantly influenced by the revised stellar mass and radius from more accurate measurements as well as the new parallax value for LHS 1140 from Gaia DR2. Furthermore, we estimate the eccentricity of LHS 1140 b to be below 0.06 with a 90\% confidence, consistent with a circular or an Earth-like orbit.

We also see strong evidence of a second terrestrial planet on a $P_c = 3.777950 \pm 0.000005$ day orbit around LHS 1140. In particular, we managed to observe several transits of LHS 1140 c with the ground-based MEarth survey as well as a single transit in April 2018 using the \emph{Spitzer Space Telescope}. The planetary signal was also discovered independently in the radial velocities from the HARPS M~dwarfs survey. Combining photometric and RV data yields a mass estimate of $m_c = 1.81 \pm 0.39$ \mearth\ and a planetary radius of $r_c = 1.282 \pm 0.024$ \rearth. The orbital eccentricity is $e_c < 0.14$ with 80\% confidence and $e_c < 0.31$ with 90\% confidence.

\begin{figure}
	\plotone{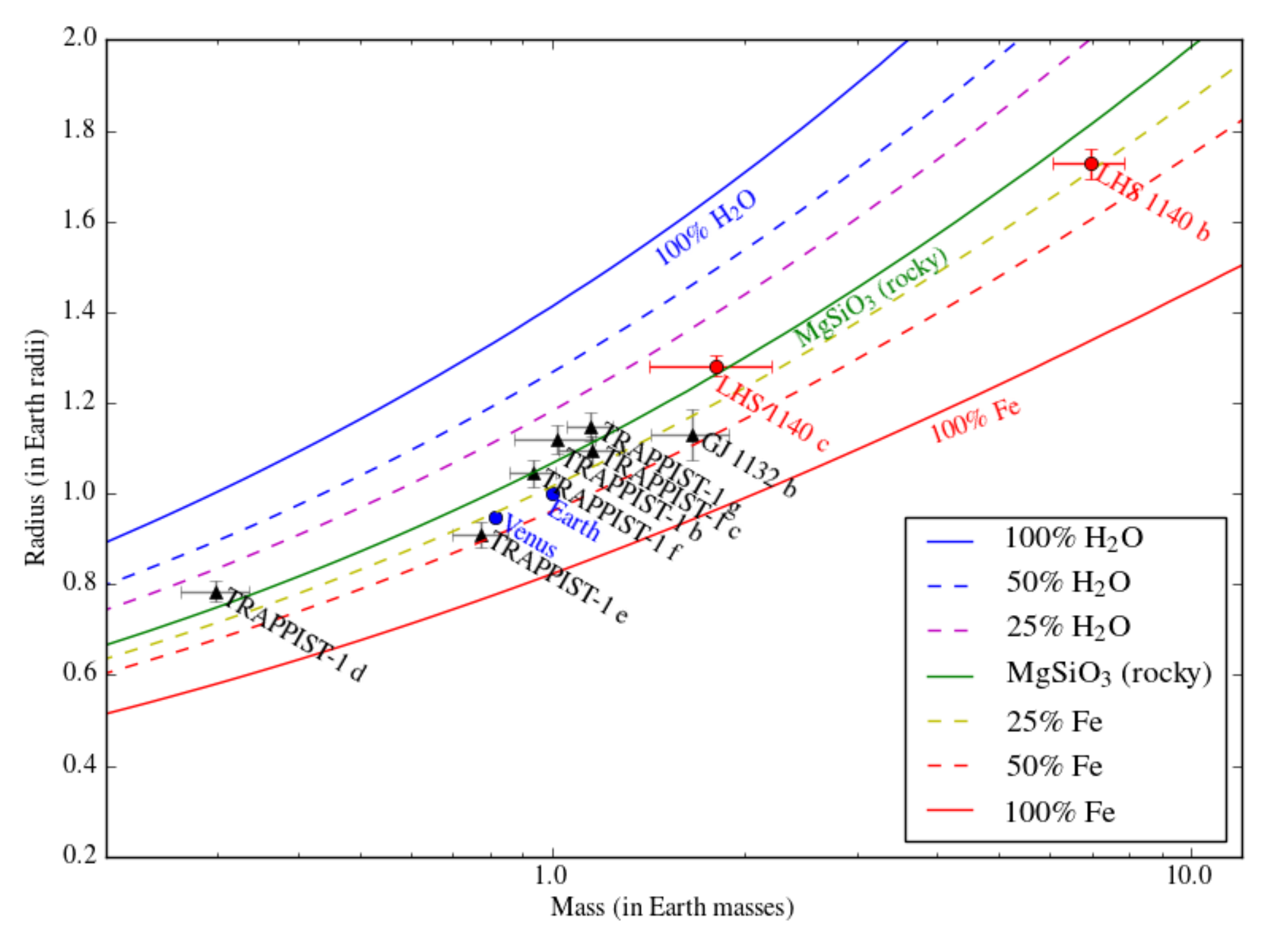}
	\caption{Mass-radius distribution of known nearby small exoplanets orbiting M dwarfs (distances less than 15 pc and host masses less than 0.3 \msun). The solid and the dashed lines show the expected mean densities for various planetary compositions \citep{Zeng2016}. The masses and radii for the TRAPPIST-1 planets were adopted from \citet{Grimm2018}. LHS 1140 b and c are both consistent with a rocky (predominantly iron and magnesium silicates) composition.\label{fig:mrrelation}}
\end{figure}

Based on our mass and radius estimates for LHS 1140 b and c, we can calculate the mean densities of the two planets: $\rho_b = 7.5 \pm 1.0$ g cm$^{-3}$ and $\rho_c = 4.7 \pm 1.1$ g cm$^{-3}$. We note that the density for planet b decreased from the $\rho_b = 12.5 \pm 3.4$ g cm$^{-3}$ reported by \citet{Dittmann2017}, mostly due to the 21\% increase in the planetary radius. Both planets are consistent with a rocky composition, as can be seen from the mass-radius diagram in Figure \ref{fig:mrrelation} which also illustrates the expected densities for several theoretical bulk compositions \citep{Zeng2016}. The density of LHS 1140 c is also consistent with the findings of \citet{Rogers2015} and \citet{Dressing2015b} who observed that planets with radii below $\sim$1.6 \mearth\ are preferentially rocky (that is, heavy enough to be composed of iron and silicates alone) whereas larger planets tend to have lower densities which implies voluminous layers of volatiles (H/He and astrophysical ices). LHS 1140 b, on the other hand, is consistent with a rocky composition despite having a radius of 1.73 $\pm$ 0.03 \rearth. Therefore, we see further evidence that the transition between planets with and without volatile envelopes is gradual \citep{Rogers2015}. We note that LHS 1140 b falls at the minimum of the occurrence rate versus planet radius recently presented by \citet{Fulton2017}.

We can also estimate the equilibrium black-body temperatures of the two planets by using the stellar temperature and Stefan-Boltzmann law. In particular, we obtain $T_{\rm eq,b}$ = 235 $\pm$ 5 K and $T_{\rm eq,c}$ = 438 $\pm$ 9 K for planets b and c, respectively, assuming a Bond albedo of 0. An Earth-like albedo of 0.3 would yield temperatures of 215 K and 401 K, respectively. The incident fluxes on the two planets compared to Earth are $S_b = 0.503 \pm 0.030$ \searth\ and $S_c = 6.16 \pm 0.37$ \searth. This implies that LHS 1140 c is hot enough to be a Mercury analogue and is likely not habitable in the Earth-based sense. It is also likely to be tidally locked and, barring gravitational perturbations from additional planetary companions, is expected to be in spin-orbit resonance. The outer planet (LHS 1140 b), however, receives half the flux incident on Earth, and could thereby be more accurately characterized as a Super-Mars. Depending on the atmospheric properties, the conservative inner and outer habitable zone limits for LHS 1140 would be 0.07 AU and 0.14 AU, respectively \citep{Kopparapu2013}, placing LHS 1140 b well within the conservative habitable zone. This is mostly consistent with the findings of \citet{Kane2018}, although we place LHS 1140 b deeper within the habitable zone due to our revised values for the orbital semi-major axis as well as the effective stellar temperature.

We note that the orbital period ratio of 1:6.55 between the two planets means that they are unlikely to be in direct orbital resonance. Orbital resonances often indicate a history of planetary migration, such as in the four-planet Kepler-223 system \citep{Mills2016} or around TRAPPIST-1 \citep{Tamayo2017}. However, recent simulations suggest that most hot super-Earths like LHS 1140 c are not expected to form in situ \citep{Cossou2014,Ogihara2015}, although resonant chains can also be disrupted by the dispersal of the gaseous disk \citep{Cossou2014}. Nonetheless, the relatively large difference between the orbital periods increases the likelihood of additional unseen companions, especially since many known planetary systems with multiple close-in super-Earths are much more compact. This view is also supported by the \textit{Kepler dichotomy} \citep{Lissauer2011,Ballard2016}. In particular, M dwarfs seem to exhibit either a single transiting planet (explained by a single planet or a multiple-planet system with large mutual orbital inclinations) or a large number of near-coplanar transiting planets \citep{Ballard2016}. The similar orbital inclinations of LHS 1140 b and c (89.89 and 89.92 degrees, respectively) suggest that LHS 1140 might belong to the latter group, with more planets yet to be discovered. Despite the fact that we do not currently see additional significant signals in the photometry and the radial velocities (as illustrated by the GLS periodogram in Figure \ref{fig:periodogram}), we strongly encourage further photometric monitoring as well as high-precision RV follow-up. We also note that LHS 1140 is one of the few nearby systems where the masses and radii of the planets can be determined with a relatively high accuracy, due to our comparatively good knowledge of the stellar parameters.

Planets in the LHS 1140 system are also potential targets for atmospheric spectroscopy observations with the \emph{James Webb Space Telescope} (JWST). \citet{Morley2017} modeled the emission and transmission spectra of LHS 1140 b for varying surface pressures and atmospheric compositions, and noted that a 5$\sigma$ detection of transmission spectral features would likely be very challenging, if the atmosphere had a similar mean molecular weight compared to that of the Earth. However, both emission and transmission spectra are sensitive to the molecular composition of the atmosphere, which can strongly depend on temperature; the emission spectra are directly sensitive to temperature as well. Due to its much higher equilibrium temperature and lower surface gravity, LHS 1140 c would be a significantly better candidate for the detection of both emission and transmission spectral features. In particular, the two planets tested by \citet{Morley2017} that have radii and temperatures similar to those of LHS 1140 c - namely TRAPPIST-1 b and GJ 1132 b - would likely require less than a dozen transits or eclipses to detect a Venus-like atmosphere, depending on the Bond albedo and surface pressure. Comparative planetology studies are often challenging due to the fact that planets in different systems evolve in very different stellar environments. Therefore, characterizing the atmospheres of multiple planets in the same planetary system would likely yield a much better understanding of how planets and their surface environments form and develop.

\acknowledgments
The MEarth team acknowledges funding from the David and Lucile Packard Fellowship for Science and Engineering (awarded to D.C.). This material is based on work supported by the National Science Foundation under grants AST-0807690, AST-1109468, AST-1004488 (Alan T. Waterman Award) and AST-1616624. This publication was made possible through the support of a grant from the John Templeton Foundation. The opinions expressed in this publication are those of the authors and do not necessarily reflect the views of the John Templeton Foundation. This work was performed in part under contract with the Jet Propulsion Laboratory (JPL) funded by NASA through the Sagan Fellowship Program executed by the NASA Exoplanet Science Institute (in support of R.D.H.). J.A.D. acknowledges support by the Heising-Simons Foundation as a 51 Pegasi b postdoctoral fellow. N.A-D. acknowledges support from FONDECYT 3180063. X.B., J.M-A., and A.W. acknowledge funding from the European Research Council under the ERC Grant Agreement n. 337591-ExTrA. R.C. acknowledges support from the Natural Sciences and Engineering Research Council of Canada. E.R.N. is supported by an NSF Astronomy and Astrophysics Postdoctoral Fellowship under award AST-1602597. N.S.C. acknowledges support from FEDER - Fundo Europeu de Desenvolvimento Regional funds through the COMPETE 2020 - Programa Operacional Competitividade e Internacionaliza\c{c}\~{a}o (POCI), and Portuguese funds through FCT - Funda\c{c}\~{a}o para a Ci\^{e}ncia e a Tecnologia in the framework of the project POCI-01-0145-FEDER-028953 and POCI-01-0145-FEDER-032113. We further acknowledge the support from FCT through national funds and by FEDER through COMPETE2020 in the form of grant UID/FIS/04434/2013 \& POCI-01-0145-FEDER-007672. This publication makes use of data products from the Two Micron All Sky Survey (2MASS), which is a joint project of the University of Massachusetts and the Infrared Processing and Analysis Center/California Institute of Technology, funded by NASA and the National Science Foundation. This publication makes use of data products from the Wide-field Infrared Survey Explorer, which is a joint project of the University of California, Los Angeles, and the JPL/California Institute of Technology, funded by NASA. This research has made extensive use of the NASA Astrophysics Data System (ADS), and the SIMBAD database, operated at CDS, Strasbourg, France.

\software{batman \citep{batman}, HARPS Data Reduction Software \citep{Lovis2007}, emcee \citep{emcee}}

\bibliography{LHS1140c.bib}

\end{document}